# Transferring Data from a Voronoi Mesh to an Adaptive Cartesian Grid in Pursuit of Self-consistent Top-down Star Formation

Sean C. Lewis[1], Brooke Polak[2], Mordecai-Mark Mac Low[1,2], Stephen L. W. McMillan[1],
Claude Cournoyer-Cloutier[3,4], Hui Li[5,6], Maite J. C. Wilhelm[7], and Simon Portegies Zwart[7]
[1] Department of Physics, Drexel University, Philadelphia, PA, USA
[2] Department of Astrophysics, American Museum of Natural History, New York, NY, USA; bpolak@amnh.org
[3] Max Planck Institute for Astrophysics, Garching, Germany
[4] Department of Physics & Astronomy, McMaster University, Hamilton, ON, Canada
[5] Department of Astronomy, Tsinghua University, Beijing, People's Republic of China
[6] Department of Astronomy, Columbia University, New York, NY, USA
[7] Leiden Observatory, Leiden University, Leiden, The Netherlands

## Abstract

Unstructured Voronoi mesh simulations offer many advantages for simulating self-gravitating gas dynamics on galactic scales. Adaptive mesh refinement (AMR) can be a powerful tool for simulating the details of star cluster formation and gas dispersal by stellar feedback. Zooming in from galactic to local scales using the star cluster formation simulation package `Torch` requires transferring simulation data from one scale to the other. Therefore, we introduce `VorAMR`, a novel computational tool that interpolates data from an unstructured Voronoi mesh to an AMR Cartesian grid. `VorAMR` is integrated into the `Torch` package, which integrates the `FLASH` AMR magnetohydrodynamics code into the Astrophysical Multipurpose Software Environment. `VorAMR` interpolates data from an `AREPO` simulation to a `FLASH` AMR grid using a nearest-neighbor particle scheme, which can then be evolved within the `Torch` package, representing the first ever transfer of data from a Voronoi mesh to an AMR Cartesian grid. Interpolation from one numerical representation to another results in an error of a few percent in global mass and energy conservation, which could be reduced with higher-order interpolation of the Voronoi cells. We show that the postinterpolation `Torch` simulation evolves without numerical abnormalities. A preliminary `Torch` simulation is evolved for 3.22 Myr and compared to the original `AREPO` simulation over the same time period. We observe similarly distributed star cluster formation between the two simulations. More compact clusters are produced in the `Torch` simulation as well as 2.3 times as much stellar material as in `AREPO`, likely due to the differences in resolution.

*Unified Astronomy Thesaurus concepts:* Computational methods (1965); Star clusters (1567); Molecular clouds (1072); Giant molecular clouds (653); Star forming regions (1565)

## 1. Introduction

Star cluster formation involves a wide range of physical processes including magnetohydrodynamics (MHD), stellar dynamics, stellar evolution, and stellar feedback. The millions of years required for star clusters to form and their early stages remaining deeply embedded in gas and dust have made it difficult to rely solely on observations to study the star cluster formation process. Numerical techniques and simulations are critical to furthering our understanding of the formation and early evolution of star clusters. Reviews in the field include M.-M. Mac Low & R. S. Klessen (2004), C. F. McKee & E. C. Ostriker (2007), S. F. Portegies Zwart et al. (2010), J. E. Dale (2015), R. S. Klessen & S. C. O. Glover (2016), M. R. Krumholz et al. (2019), P. Girichidis et al. (2020), and M. G. H. Krause et al. (2020).

Many different types of simulation software are used to model the formation of star clusters. These simulations can differ from one another at a fundamental level, using entirely different data structures and algorithms to accomplish their numerical tasks. Two simulation tools, `AREPO` (V. Springel 2010; R. Weinberger et al. 2020) and `Torch` (J. E. Wall et al. 2019, 2020), are pertinent to this work. `AREPO` uses a Godunov solver on a Voronoi mesh to model gas hydrodynamics while following the gravitational potential with the tree-based Poisson solver from `GADGET-2` (V. Springel 2005). The version we use includes stellar radiative and mechanical feedback from the star formation module `SMUGGLE` (F. Marinacci et al. 2019; H. Li et al. 2020). The `SMUGGLE` model does not inject explicit wind particles. `Torch` consists of an interface between the stellar dynamics and stellar evolution codes in `AMUSE` (S. Portegies Zwart et al. 2009, 2013; F. I. Pelupessy et al. 2013; S. Portegies Zwart & S. McMillan 2018) and the adaptive mesh refinement (AMR) MHD code `FLASH` (B. Fryxell et al. 2000; A. Dubey et al. 2014), in which we include ray-tracing radiative transfer (C. Baczynski et al. 2015). `Torch` simulations are at the cluster scale and follow the collisional stellar dynamics with star particles down to $0.08\,M_\odot$, while `AREPO` is used at galactic scales and uses gravitational softening of star particles with masses representing hundreds to thousands of stars. `AREPO` and `Torch` both provide unique insight to aspects of the star formation process, but individually they cannot accurately model the entire process. Data produced by one tool are usually incompatible with other tools, preventing researchers from easily using the output data from one tool as input data for another with different physics and scales.

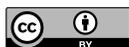

One recent effort transferred data from a large-scale cosmological model run using smoothed particle hydrodynamics (SPH) to models of individual magnetized galaxies using an AMR code (A. Stacy et al. 2022). A tool for converting data from FLASH to GADGET-2 was developed (J. Karam & A. Sills 2024) and used to model mergers between embedded clusters (C. Cournoyer-Cloutier et al. 2024a). Neither method treats a Voronoi mesh.

Another set of past efforts has bridged the gap between simulation software suites for visualization purposes. Notably, the yt project (M. J. Turk et al. 2011) approximates the Voronoi mesh voxels in AREPO as smoothed particles using an SPH kernel to map the simulation data to Cartesian grids. However, yt was originally designed to process large AMR grids from software like ENZO (G. L. Bryan et al. 2014), FLASH (B. Fryxell et al. 2000), and ATHENA (J. M. Stone et al. 2008). Similarly, POLARIS (S. Reissl et al. 2016, 2019) has the ability to process and visualize data from software suites of varying data structuring methods and geometries. While methods such as these are useful and even critical for visualization purposes, they do not allow for the transfer of data between different simulation tools to continue the evolution of a simulation. This is because POLARIS does not output a grid but rather a simulated observation. In the case of yt, Voronoi cells are treated as SPH particles, so the well-defined cell values are averaged over to make an SPH kernel resulting in lost resolution. AREPO gives second-order Godunov values, so the averaging into SPH particles to get local values results in shocks being broadened and density peaks softened. Preserving these sharp gradients given by the Godunov solver has many advantages. For example, for applications in radiative transfer or interstellar medium physics a hot and cold cell would be averaged to produce an unphysically warm cell, whereas the temperatures would be preserved in a grid interpolation.

The ability to start a star formation simulation in one software suite and continue the simulation in another is particularly useful for star cluster formation researchers as it provides a novel method for creating robust simulations with a desirable quality: self-consistent initial conditions. The initial conditions of star cluster formation at the scale of individual clouds are sufficiently complex to prevent the formulation of a perfectly accurate numerical description. In response, models that focus on this parsec-scale dynamical range have employed several techniques to approximate self-consistent initial conditions. Historically, a turbulent sphere of cold, dense gas has been a common approach (e.g., M. R. Bate et al. 1995, 2002; J. E. Dale et al. 2012, 2014; J. E. Wall et al. 2019, 2020). This approach has the benefit of being easy to describe mathematically and easy to generate. Similar isolated initialization efforts include colliding cylindrical clouds to simulate the collision of gas flows and filaments (e.g., E. Vázquez-Semadeni et al. 2017; C. L. Dobbs et al. 2020, 2022) or stirred, periodic boxes in which filamentary structures emerge from an initially uniform turbulent medium (e.g., D. A. Tilley & R. E. Pudritz 2004; M. R. Krumholz et al. 2011; R. Pillsworth & R. E. Pudritz 2024). Ultimately, simulations that use these isolated initial conditions will lack realistic giant molecular cloud (GMC) initial conditions but excel in simulating the dynamics of and feedback from individual stars. At the same time, large-scale galactic simulations are able to form more realistic GMCs within a dynamic galactic environment (S. M. R. Jeffreson et al. 2022; H. He et al. 2023; B. Zhao et al. 2024). With VorAMR, GMCs that form under realistic dynamic galactic conditions can then be used as initial conditions for star-forming environments in the Torch package.

We present VorAMR, a computational tool built in AMUSE, which allows for the transferring of data from AREPO simulations to Torch by interpolating Voronoi mesh data onto an AMR grid. It is recommended to convert Voronoi cells to triangular meshes to minimize interpolation error (Q. Du & M. Gunzburger 2002). However, most currently used astrophysics codes use an adaptive Cartesian mesh. While there are tools that use Voronoi cells to generate two-dimensional (2D) unstructured meshes (e.g., C. Talischi et al. 2012) and 2D triangular meshes (e.g., J. Koko 2015), there were no existing tools for porting to a three-dimensional (3D) adaptive Cartesian mesh before VorAMR.

In this work, we explain how VorAMR translates data from AREPO to Torch and perform such a translation to use a GMC complex from a zoom-in galaxy-scale AREPO simulation (H. Li et al. 2020) as initial cloud conditions in Torch. In Section 2 we describe the computational structure of VorAMR, the nature of the AREPO source data, and notable capabilities and features of VorAMR. In Section 3 we note the AMR grid maximum refinement level dependency on the accuracy of the VorAMR interpolation. We also showcase a proof-of-concept star cluster formation simulation. Lastly, in Section 4 we discuss the impact of VorAMR on computational astrophysics and future VorAMR development efforts.

## 2. Method

### 2.1. VorAMR Structure and Logic

VorAMR consists of two communication pipelines between AREPO data and AMUSE and FLASH to interpolate AREPO data into FLASH. From here on, we refer to the input AREPO data as source data, mesh, mesh-generating particles, or mesh elements and the postinterpolation data within FLASH as the *grid*, which is comprised of *blocks* that themselves contain by default $16^3$ cells. Refinement is done at the block rather than cell level in FLASH.

The AREPO mesh–generating particles are treated as special particles within FLASH. The particle distribution is used to set the initial refinement criteria to build the AMR grid, after which the particles are discarded. FLASH refines on particle count, which was implemented in A. Dubey et al. (2012). At the same time, AMUSE uses the same AREPO gas mesh particle data to build a kdtree. The kdtree partitions the AREPO mesh data into a searchable 3D data structure. The kdtree is then able to rapidly determine which AREPO mesh element is closest to any given a point in space. Using this, each FLASH grid cell has the nearest AREPO gas field value assigned to it such as density, internal energy, and velocity. Once FLASH has finished building the refined grid, AMUSE requests the data structure of each FLASH block. FLASH passes each empty block to AMUSE where the field values of the nearest mesh particle to each FLASH cell within the block are found via the kdtree and assigned to each cell. Using a nearest-neighbor interpolation has some inherent biases, which we discuss in Section 3. After the interpolation is complete, the populated block is returned to FLASH.

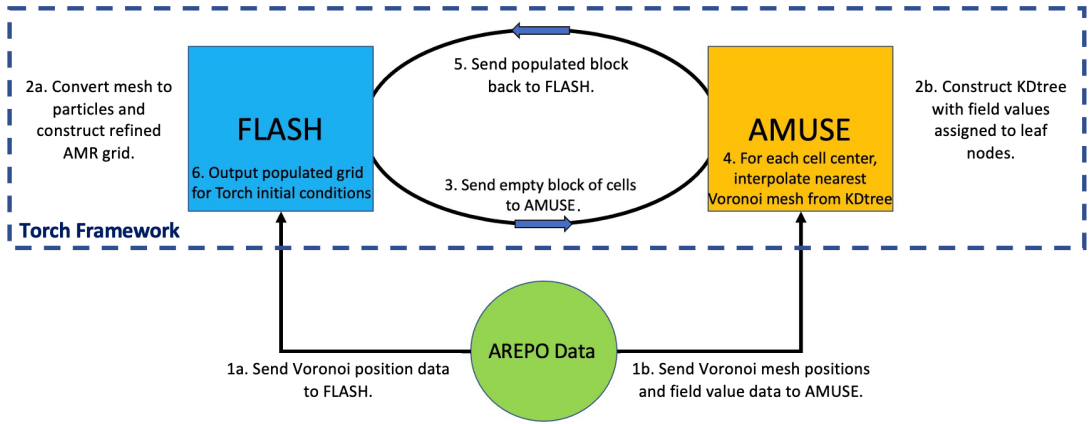

**Figure 1.** `VorAMR` data flow (see Section 2.1). `VorAMR` builds upon the existing `Torch` framework in `AMUSE` and allows `AREPO` data to be read in and processed into `FLASH`-compatible data.

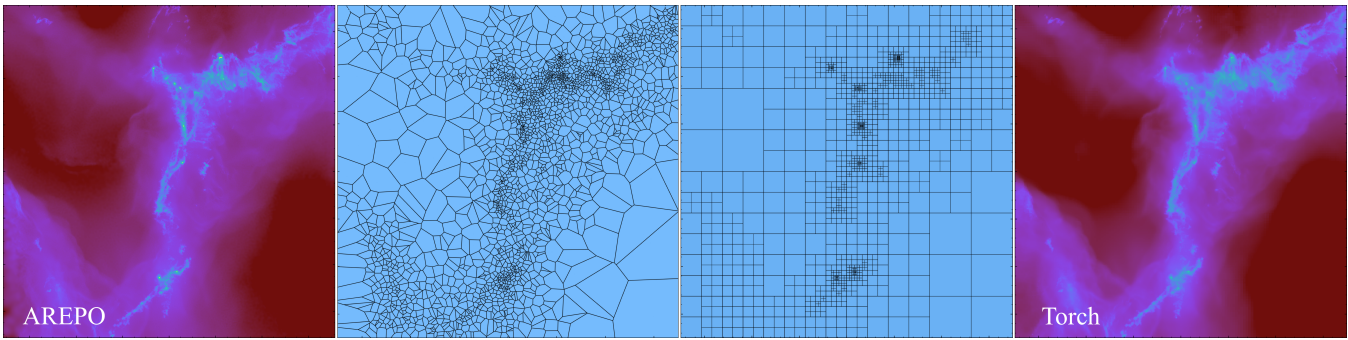

**Figure 2.** Far left: gas density projection of `AREPO` source data. Middle left: Voronoi mesh generated with the scipy.voronoi library from every 4096th data point of `AREPO` source data (for visualization purposes). Middle right: empty `FLASH` AMR grid generated from all `AREPO` data with AMR blocks annotated and level 12 maximum refinement. Far right: populated `FLASH` AMR grid after `VorAMR` interpolation showing projected gas density; initial conditions of a `Torch` simulation. The domain is 1.43 kpc wide.

The `VorAMR` logic flow is shown in Figure 1 can be summarized as follows:

1. (a) Voronoi mesh–generating particle coordinates are sent to `FLASH`.
   (b) Coordinates and field value data are sent to `AMUSE`.
2. (a) The coordinates are represented as particles in `FLASH`, which then constructs an empty AMR grid that satisfies the criterion that every block should contain no more than $16^3$ particles.
   (b) `AMUSE` constructs a `kdtree` from the particles with source data field values assigned to leaf nodes.
3. `FLASH` sends `AMUSE` an empty $16^3$ matrix, representing a single `FLASH` grid block.
4. `AMUSE` then passes the block into an *N*-dimensional nearest-neighbor interpolation routine that interpolates field data corresponding to the closest `kdtree` leaf for each of the $16^3$ block cells, creating a four-dimensional matrix of matrices.
5. `AMUSE` sends the populated block matrix back to `FLASH` via the interface, mapping each field value to the cells within the original `FLASH` block, populating the AMR grid structure. ⇌ Repeat steps 4–6 for each `FLASH` grid block.
6. `FLASH` can then immediately output an HDF5 checkpoint file or launch straight into a `Torch` simulation.

During refinement, each `FLASH` block is flagged to refine or derefine until the contained-particle criterion is satisfied. The blocks of the AMR grid must also adhere to the restriction that neighboring blocks differ by no more than a single refinement level. This results in some regions of the computational domain being forced to overrefine such that they oversample low-density Voronoi cells. This method also does not guarantee that high-refinement Voronoi regions are interpolated into single `FLASH` grid cells as a single cell may contain multiple Voronoi regions if the particles in a block are concentrated. This does lead to a small amount of data loss (<5%). We discuss the extent of this interpolation error in Section 3.

If the source data are not centered at (0,0,0), `VorAMR` detects the dimensions of the computational domain of the source data and performs a coordinate shift so that the data are centered at the origin. A similar centering is applied to the gas velocity data: the center-of-mass velocity of all gas in the domain is calculated and subtracted from the velocity field. This is a critical step as the source data can still include bulk gas velocities representing the orbit of gas around the galaxy, which we do not model postinterpolation. `VorAMR` also accounts for unit changes between `AREPO`, which uses cosmological units, and `FLASH`, which uses centimeter-gram-second units.

Figure 2 is an illustrative example of the stages of the `VorAMR` pipeline. Data are first extracted from the source file,

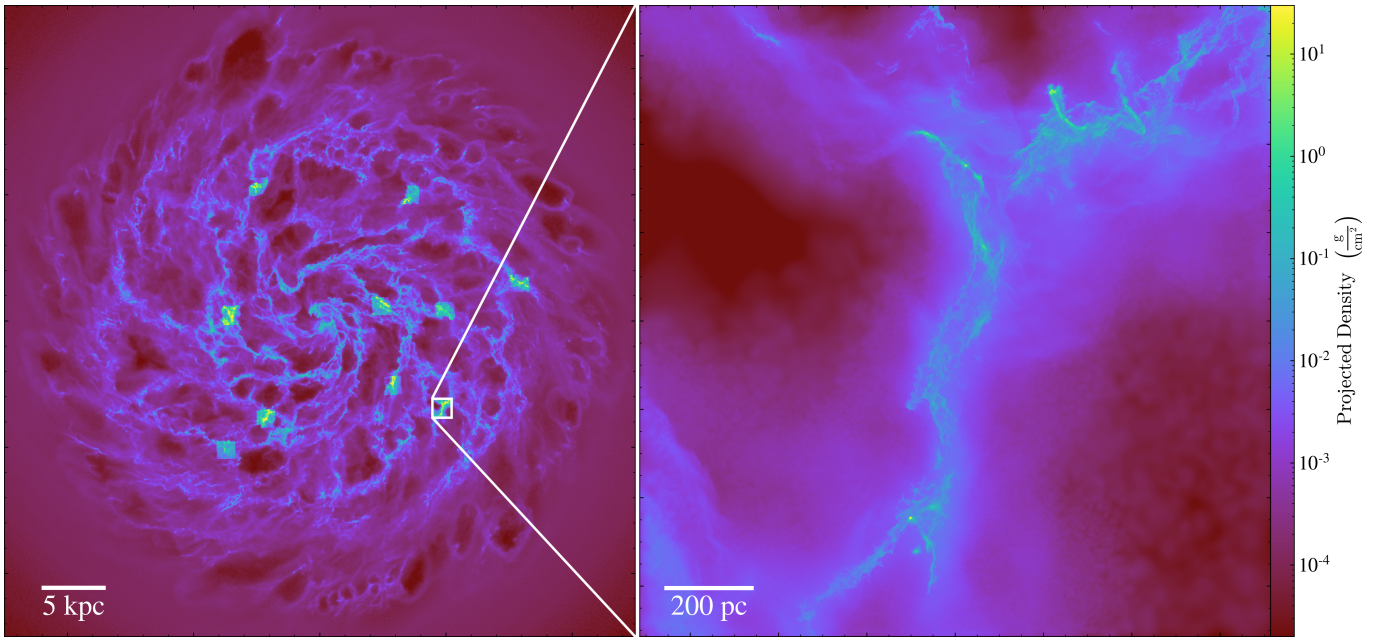

**Figure 3.** Left: the full `AREPO` galaxy simulation from which the source data in this work are sampled. This galaxy is the SFE10 model detailed in H. Li et al. (2020). The squares of high resolution are the various zoom-in regions modeled. Right: the zoom-in region used as source data for this work.

in this case an `AREPO` simulation of a collapsing GMC complex. Then, the structure of the source mesh is used to construct a `FLASH` grid that mimics the resolution of the source mesh by considering each mesh-generating particle as an object on which to refine. The empty grid is populated with source data with the nearest-neighbor interpolation scheme, and finally the grid is output as a `FLASH` checkpoint file.

The checkpoint file produced by `VorAMR` can be read by other `Torch` versions to include physics or data-tracking not native to the original source data, such as binary star formation (C. Cournoyer-Cloutier et al. 2021, 2023, 2024b), controlled massive star formation (S. C. Lewis et al. 2023), protoplanetary disks (M. J. C. Wilhelm et al. 2023), or stellar jets (S. M. Appel et al. 2025).

### 2.2. Source Data

To illustrate the application of `VorAMR` to scientific data, we use one of the 40 star-forming regions simulated by a novel zoom-in technique (H. Li et al. 2025, in preparation) from galaxy-scale simulations (H. Li et al. 2020). GMCs of interest were chosen for zoom-in due to their future high star formation rates. High-resolution zoom-ins of these GMCs were then run in `AREPO` (see right panel in Figure 3). The star formation complex detailed in this work is taken ∼8 kpc from the galaxy center. The zoom-in domain is ∼1 kpc per side. The total gas mass within the domain is $1.81 \times 10^7 \, M_\odot$. Just before data transfer, 1% of this gas has a temperature < 100 K, which is the regime that will most likely form stars once evolved in `Torch`.

### 2.3. Other Features

#### 2.3.1. Background Potential

Often, astrophysical simulations include gravitational potentials from background influences, such as dark matter. A consistent transfer of data between numerical tools must include such background potentials. The background-accel module in `VorAMR` calculates the functional form of a static background gravitational acceleration.

In the case of the `AREPO` snapshots showcased here, there is an analytic dark matter potential and softened massive particles that represent different components of the stellar galactic gravitational potential. The particles represent a significant fraction of the mass present in the simulation snapshots and so must be modeled to preserve the self-consistent nature of the initial conditions. We pass the background particles to a BHTree gravity solver (J. Barnes & P. Hut 1986) in `AMUSE` and sample the gravitational acceleration across the domain. We then pass those data through a median filter to smooth the influence of clumped background potential particles. We fit a one-dimensional, third-order polynomial that we use as the analytical solution to the background field in the vertical direction. The field can then be included in `Torch` by calculating and adding the position-dependent background acceleration terms to the gravitational acceleration data array that `FLASH` produces with its gravity solver. The background gravity field remains constant throughout the simulation. The gas self-gravity is calculated by `FLASH`.

#### 2.3.2. Custom Zoom-in

`VorAMR` can zoom in on user-specified regions of the given input data, as seen in Figure 4. This allows for higher refinement for regions of interest, as well as reduced computation time for the initialization of the grid. In this sense, `VorAMR` allows for a continuation of the GMC identification and zoom-in described by H. Li et al. (2020). To reduce the computational requirements, we mask out all input data that are not contained within the zoomed-in computational domain. The velocities of the interpolated gas are shifted to be relative to the gas center-of-mass velocity of the zoomed-in region. This helps ensure that the zoomed-in gas does not quickly exit the computational domain. One consideration, however, is that once a zoom-in simulation is initialized, gas data outside of the subregion are discarded and do not gravitationally affect and cannot enter the zoomed-in domain.

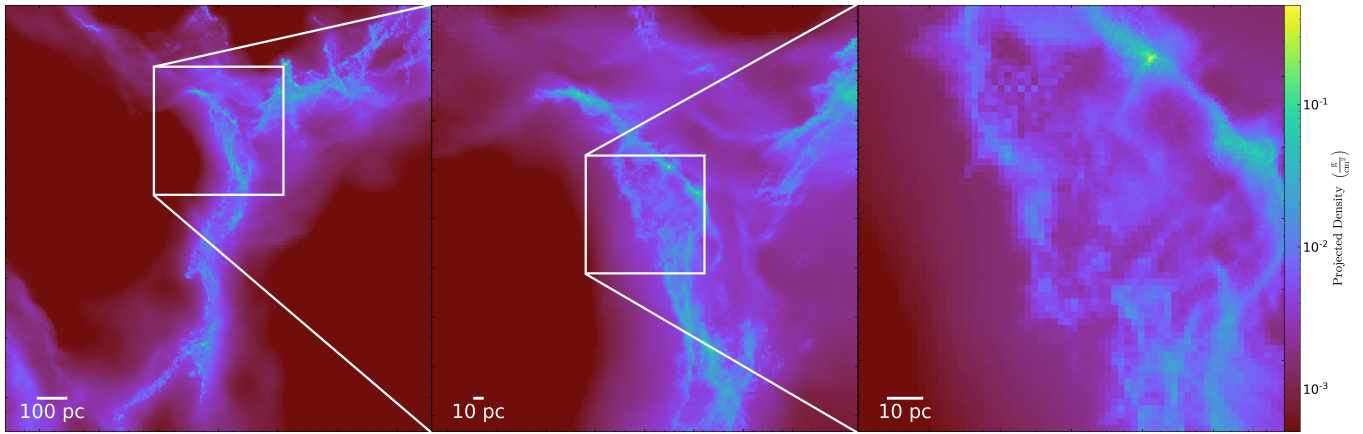

**Figure 4.** Demonstration of `VorAMR`'s zoom-in capability. Any user-defined cubic region within the full computational domain (left) can be zoomed in to and evolved as a normal `Torch` simulation. Here, two zoom-in regions are shown: a 300% zoom (center) and a 700% zoom (right). All panels show projected gas density and have a maximum refinement level of 5, corresponding to a resolution of 5.6, 1.87, and 0.80 pc from left to right. Both zoom-ins are centered at the same point within the original computational domain to show the effect of enhanced detail while maintaining the same refinement level and criteria.

Zoom-ins must also account for the shift in background potential such that the potential of the zoom-in region is now at the origin. We accomplish this by shifting the analytical background potential by the zoom-in center offset in $z$.

### 2.3.3. Localized Refinement

An alternative to using `VorAMR` to zoom in on regions of source data is to initially refine on user-specified regions of the given input data while forcing the rest of the domain to remain derefined. This solves two problems with zoom-in regions: (1) the resulting `Torch` evolution is cut off from the dynamical influence of the rest of the original source domain and does not experience tidal effects of gas structures from beyond the zoom-in region, and (2) surrounding gas can no longer flow into the zoomed-in region, which after a fraction of a crossing time becomes problematic.

The user-specified refinement capability of `VorAMR` can use all the capabilities of the PARAMESH AMR library (P. MacNeice et al. 2000) that forms the basis of the AMR in `FLASH`. Focused refinement results in a significantly decreased computation time for initial grid construction and so is also an ideal method for visualizing subregions of an input domain. Figure 5 shows two grids built using the same Voronoi source data, one allowing for refinement throughout the entire AMR domain, the other restricting refinement within a spherical region with a radius of 75 pc.

The localized refinement is accomplished by only passing the Voronoi source data within the refinement region to FLASH so that a grid is only generated by FLASH within the refinement region. This way, the grid-building routines of `VorAMR` have no source data to refine on outside of the user-defined region, restricting the refined portions of the domain. However, the kdtree from which data are interpolated is still built to contain source data from the entire domain. Therefore, `VorAMR` is still able to interpolate data across the entire domain regardless of whether the `FLASH` blocks lie within or outside the refinement region. Blocks outside the region of localized refinement will attempt to derefine to the minimum allowed refinement level. Note that PARAMESH requires there to be a maximum refinement level difference of 1 between neighboring blocks, and so blocks immediately outside of the localized refinement region will be in some intermediate refined state.

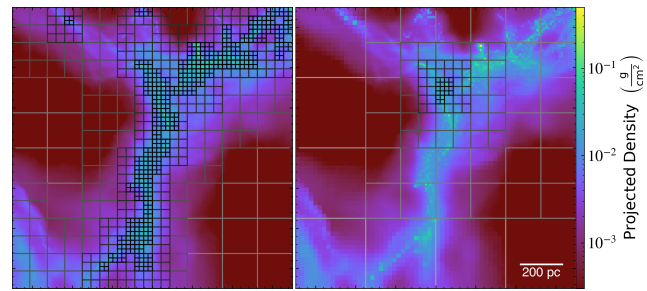

**Figure 5.** Comparison between the AMR structure using the full input data set (left) and localized refinement parameter (right). Local refinement is centered at the same point as the zoom-in regions in Figure 4. The FLASH grid blocks are shown to demonstrate the reduction in total blocks necessary for a localized refinement.

## 3. Proof of Concept and Performance

Our `Torch` simulation uses sink particles (C. Federrath et al. 2010), the high-performance stellar dynamics code `PETAR` (L. Wang et al. 2020), and the stellar evolution code `SeBa` (S. F. Portegies Zwart & F. Verbunt 1996). We confirm that `Torch` initial conditions produced by `VorAMR` are well behaved when evolved. We initialize and evolve several `Torch` grids where the gas is largely bound, so we expect that it will readily form stars as it did in the original source simulation. Indeed, in these proof-of-concept simulations, sinks readily form, gas continues to collapse, star particles are placed, and stellar feedback from ionizing radiation and winds behaves normally.

We compare the evolved `Torch` simulation to the original `AREPO` data in Figure 6, and we find similarly located star-forming regions after 3.2 Myr. `AREPO` and `Torch` have entirely different methods for placing stars and introducing stellar feedback as `AREPO` particles represent whole stellar populations rather than individual stars like in `Torch`. Therefore, we do not expect the two simulations to look identical despite evolving from the same source data for the same amount of time. In addition, the preliminary `Torch` run shown here has a maximum refinement level of 5 with a resolution of 5.6 pc[8] per cell side at the highest level of refinement. The AREPO mesh has a minimum adaptive gas softening length of 0.07 pc, and the median size of the smallest cell across all snapshots is 0.1 pc. The Torch run forms more compact clusters and 2.3 times as much stellar mass, $1.81 \times 10^5\,M_\odot$ as compared to $7.87 \times 10^4\,M_\odot$ in the AREPO run. We will investigate these runs further in future publications.

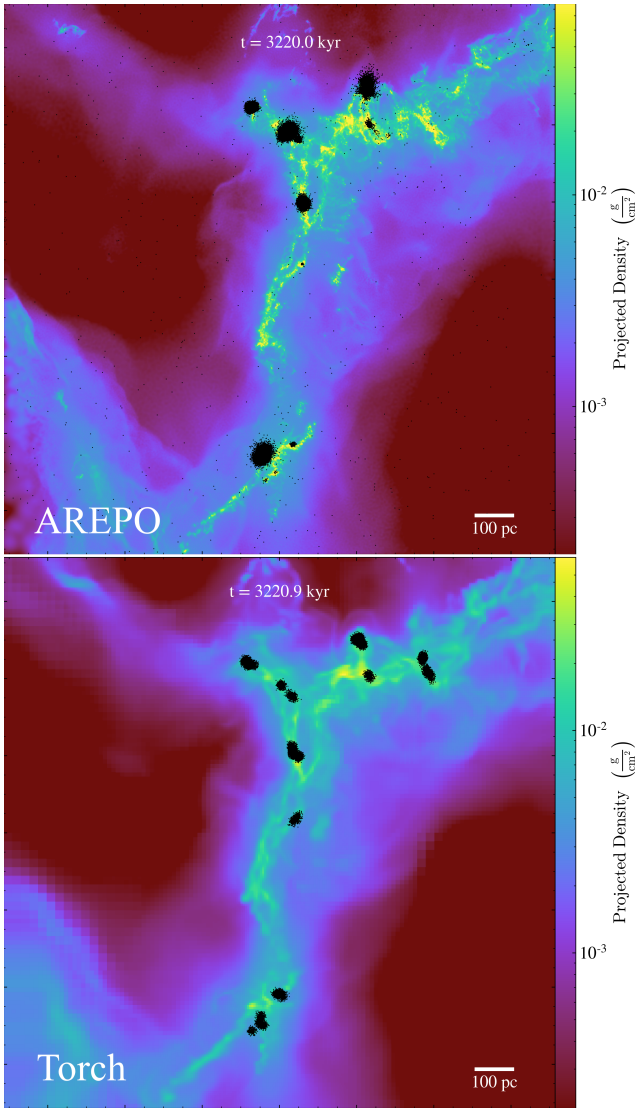

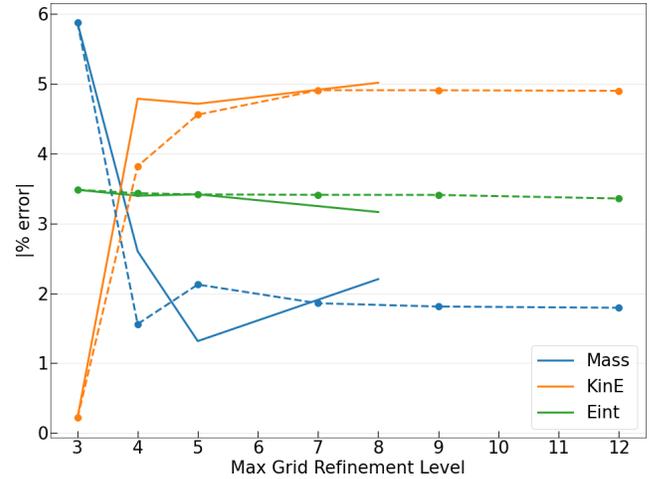

**Figure 6.** Gas column density at 3.22 Myr beyond the transfer time of (top) the source AREPO simulation and (bottom) the Torch simulation. The black points are projected star particles that have formed from the gas. Note that the color bars have different ranges.

The nearest-neighbor interpolation of the source data into the FLASH grid naturally results in errors. For example, if a single FLASH cell bounds a region occupied by multiple source data points, only the data closest to the FLASH cell center are brought onto the grid. This also means the interpolation is biased toward larger volume mesh elements. In such an undersampling case, caused either by tightly clumped source data or an insufficient maximum grid refinement level, some source data are lost. In general, undersampling is not anticipated to be a large effect as any dropped data will typically be similar to the data that are interpolated onto the FLASH grid. Furthermore, in production runs where the FLASH resolution will match the AREPO resolution, the interpolation error in regions of interest where star formation occurs will be small. Most of the interpolation error will occur in the large low-density regions outside the dense cores that remain at a low refinement level. Using a higher-order interpolation would lower this interpolation error in low-density regions, but this would not improve the simulation results because the low-density large-volume regions are inconsequential to star formation. These regions exist only to provide inflow/outflow to the star-forming filaments.

---

[8] The proof-of-concept run was done with low resolution for computational efficiency. Production simulations will be at higher resolutions ($\lesssim 0.1$ pc).

**Figure 7.** Error in total mass, kinetic energy, and internal energy after first-order nearest-neighbor interpolation of AREPO mesh data onto an AMR FLASH grid at various levels of maximum refinement. Solid lines represent the error from FLASH grids that are allowed to adaptively refine across the entire domain. Dashed lines are errors from FLASH grids that only refine in a region of interest 75 pc in radius, representing 4.12% of the original domain volume, as seen in Figure 5.

Using the same source data discussed throughout this paper, we find that while there is some error after interpolation in total mass, kinetic energy, and internal energy, none rise above 5%, as seen in Figure 7. Intuitively, we would expect a lower error at higher levels of refinement, but that is not seen. We suspect that at low refinement level grid cells act as broad smoothing filters, smearing a single AREPO mesh element across coarse FLASH cells despite each cell often containing many mesh elements. However, source data in a slightly different orientation, say, offset by a few percent, are more likely to cause the error of the coarse grids to fluctuate as data interpolated into the cells are more likely to vary. Therefore, while a coarse grid will tend to average out any interpolation error that occurs, it is more likely to see a range of errors at low grid refinement, from very low to very high. We see this behavior in Figure 7, where the interpolation error fluctuates at lower grid refinement but becomes more stable at higher refinement as the grid structure approaches a comparable organization to the source AREPO mesh. Due to the adaptive refinement, in all test cases there remain large regions of coarse grid at the lowest refinement level. Mapping a large irregular shape to a large square will always result in some level of interpolation error. The error saturates at a nonzero value at higher refinements because the error in the coarse regions remains the same regardless of the maximum refinement level.

We compare the interpolation error for grids where the entire computational domain is permitted to refine based on the AREPO mesh structure to grids where refinement is only allowed inside a 150 pc diameter region centered on the densest AREPO mesh cell. The error does not improve monotonically as refinement increases for either the full-grid or region-of-interest tests. In addition, the full-grid and region-of-interest errors approach one another at high refinement. This indicates a source of interpolation error in low-refinement regions in addition to the high-refinement region data loss mentioned previously. The first-order nearest-neighbor interpolation method does not properly mimic the distribution of data in AREPO Voronoi cells, an effect most apparent at low refinement where many FLASH cells have the data of a single AREPO mesh cell interpolated into them. To improve the interpolation error in conserved quantities, a more precise FLASH grid refinement scheme and a higher-order interpolation method would be needed.

## 4. Discussion

### 4.1. Impact

The translation of data from a Voronoi mesh onto an AMR grid creates another avenue for increased collaboration between researchers. Simulation tools grow robust and complex over time, leading to the tendency of researchers to become deeply entwined with unique but isolated software. This can lead to difficulties in translating computational advances made in one software tool to another. VorAMR enables researchers to collaborate with each other to maximize the reach and spread of their data and computational techniques.

As a specific example of such collaboration, we show in this work how VorAMR becomes a critical linkage in the star cluster simulation pipeline, allowing researchers to initialize star cluster simulations using gas initial conditions from galactic-scale simulations. The star formation process is influenced by physics from an enormous range of length scales, from galactic potential and outflows down to the dynamics and feedback from individual stars. The range of length scales makes it prohibitively expensive for any one simulation to include all physical processes without significant approximations. For example, a simulation of an entire galaxy may include single particles that represent thousands or tens of thousands of stars, or a simulation of a single GMC may include a background galactic potential without actually simulating the galactic dynamics. Researchers may also use idealized initial conditions for the gas cloud such as stirred boxes, turbulent spheres, or colliding cylinders. VorAMR enables star formation simulation tools like Torch to access realistic GMC initial conditions from galactic simulations. VorAMR provides a better alternative to idealized initial conditions in Torch simulations.

VorAMR also provides a novel way to visualize Voronoi mesh–based hydrodynamical data. Voronoi meshes are notoriously difficult to produce visualizations from, and researchers have often relied on suboptimal approximations. A common method is to treat each Voronoi mesh cell as a particle and visualize the data using an SPH averaging kernel. This is not ideal as it smooths across the original Voronoi mesh data. Otherwise, researchers have transferred the Voronoi mesh data onto a uniform grid at the highest resolution of the model, which becomes infeasible for simulations with large dynamic ranges. The advantage of the AMR grid is that it represents the same information at much lower cost in memory. Robust tools like yt (M. J. Turk et al. 2011) are built for efficiently visualizing large AMR grids. By interpolating onto an AMR grid, VorAMR provides a form of mesh visualization without relying on the use of SPH kernel averaging.

Currently, VorAMR interpolates data from AREPO simulations into Torch, but in principle it could be expanded to allow data from any simulation suite to be translated into Torch, including data from other Voronoi mesh codes, SPH codes, or even other AMR codes.

### 4.2. Limitations

The interpolation of data from an AREPO mesh into a FLASH grid results in a persistent error in the grid field values. Some error is of course expected as a block-based AMR grid will never be able to exactly match the distribution of an unstructured mesh. We determine that the interpolation error reaches a few percent and remains consistent with increasing FLASH grid refinement. Further development to implement higher-order interpolation schemes and more precise FLASH block refinement criteria should improve the interpolation error. Besides cloud in cell, another potential avenue for higher-order interpolation would be to follow the method outlined in J. Borrow & A. J. Kelly (2021) for mapping SPH data onto a grid, perhaps using the Voronoi cell size as a proxy for the SPH kernel size.

VorAMR is closely tied to source data from AREPO simulations and so cannot yet be applied to a diverse set of input data structures. Because of the modular nature of VorAMR and Torch, future data structures such as other AMR grids or SPH simulations as sources can be accommodated.

VorAMR detects and operates on AREPO gas mesh-generating particles. VorAMR does not automatically handle other types of AREPO particles like star or wind particles. We have analytically derived an appropriate average background galactic potential based on the corresponding AREPO particle set. We have not attempted to convert AREPO star particles into Torch star particles. Therefore, the presence of AREPO star particles represents a fraction of source domain total mass that is not interpolated onto the FLASH grid via VorAMR. Neglecting such particles in the AREPO run that we use results in only a fraction of 1% to a few percent of additional error in the total mass and gravitational energy. It is good practice to start a VorAMR run before a significant amount of stars have formed to minimize this error. VorAMR was run in serial[9] for the work presented in this paper, but it is now available to run in parallel.[10]

## 5. Summary

In this work we describe the use of AMUSE to create a pipeline connecting output Voronoi mesh data from AREPO to input AMR grid data in `FLASH` to allow further evolution with the `Torch` star cluster simulation package.

---

[9] https://bitbucket.org/torch-sf/torch/src/vorch-v1.0
[10] https://bitbucket.org/torch-sf/torch/src/vorch-v2.0

By using the `AREPO` Voronoi mesh–generating particles to drive refinement, we use native `FLASH` routines to construct an empty adaptively refined grid. We currently refine until there is an average of one mesh-generating particle per cell within each `FLASH` block, though we note that this still does not guarantee that every particle is in a separate cell. Simultaneously, we construct a $k$-dimensional lookup tree where each leaf node consists of the Cartesian coordinates of the `AREPO` mesh–generating particle and the gas field values of density, internal energy, gravitational potential, and three-dimensional velocity. We then use a first-order nearest-neighbor interpolation scheme, taking advantage of the increased lookup speed of the $k$-dimensional tree, to query and fill each `FLASH` cell with gas field values from the nearest `AREPO` particle.

We find that our interpolation scheme results in an error of a few percent in the global conserved quantities of total mass and energy. Gas within postinterpolation `Torch` runs continues to collapse to form multiple star clusters in filamentary structures over several megayears—in the same regions that formed stars in the `AREPO` simulation—with no indications of numerical instability, further indicating successful interpolation for our use case.

We describe several capabilities of the `VorAMR` data pipeline that work to provide a diversified range of uses in astrophysical simulation. Subregions of source data can be selected for interpolation onto the `FLASH` grid in `Torch`, allowing for zoom-ins to regions of interest. We also provide a localized refinement module that allows subregions of the `FLASH` grid to undergo refinement while maintaining lower refinement throughout the rest of the computational domain. Lastly, we implement a background gravitational module so that additional gravitational influences from source data such as dark matter can be represented in the `Torch` simulation.

We note that the `VorAMR` data interpolation pipeline can be used for any source data so long as it can be represented as particles, though more robust interpolation methods would be needed for consideration of representations like SPH kernels.

We hope this work inspires similar efforts to build bridges between hydrodynamical simulation tools, promoting an open, interconnected scientific community.

## Data Availability

The data from the simulations and figures within this article will be shared upon reasonable request to the corresponding author. With the publication of this work, we make the `VorAMR` data interpolation pipeline public as part of the `Torch` distribution.

## Acknowledgments

The authors thank the referee for the insightful comments and questions, one of which led to an explanation of why the interpolation error saturates. The authors also thank Eric Andersson and Sabrina M. Appel for the useful discussions. This work was supported by NSF grants AST18-15461 and AST-23-07950, and NSF ACCESS grants PHY220160 and PHY240335. C.C.C. acknowledges funding from a Canada Graduate Scholarship—Doctoral from the Natural Sciences and Engineering Research Council of Canada (NSERC). H.L. is supported by the National Key R&D Program of China No. 2023YFB3002502, the National Natural Science Foundation of China under grant No. 12373006, and the China Manned Space Program through its Space Application System. We acknowledge John Zuhone for his consistent and timely assistance with `yt` visualizations and processing.

*Facilities:* Snellius; SURF—Dutch National Supercomputing Center, Stampede2 and Stampede3; Texas Advanced Computing Center.

*Software*: `Torch` (J. E. Wall et al. 2019, 2020), `AMUSE` (S. Portegies Zwart et al. 2009, 2013; F. I. Pelupessy et al. 2013; S. Portegies Zwart & S. McMillan 2018), `FLASH` (B. Fryxell et al. 2000), `yt` (M. J. Turk et al. 2011), numpy (T. E. Oliphant 2007), scikit-learn (F. Pedregosa et al. 2011), matplotlib (J. D. Hunter 2007), HDF5 (S. Koranne 2011).

### ORCID iDs

Sean C. Lewis ⓘ https://orcid.org/0000-0003-4866-9136
Brooke Polak ⓘ https://orcid.org/0000-0001-5972-137X
Mordecai-Mark Mac Low ⓘ https://orcid.org/0000-0003-0064-4060
Stephen L. W. McMillan ⓘ https://orcid.org/0000-0001-9104-9675
Claude Cournoyer-Cloutier ⓘ https://orcid.org/0000-0002-6116-1014
Hui Li ⓘ https://orcid.org/0000-0002-1253-2763
Maite J. C. Wilhelm ⓘ https://orcid.org/0000-0002-3001-9461
Simon Portegies Zwart ⓘ https://orcid.org/0000-0001-5839-0302